\documentclass[12pt,letterpaper]{article}
\usepackage[margin=1in]{geometry}
\usepackage[english]{babel}
\usepackage[utf8x]{inputenc}
\usepackage[T1]{fontenc}

\usepackage[numbers]{natbib}

\usepackage{amsmath}
\usepackage{amssymb}
\usepackage{graphicx}
\usepackage{adjustbox}
\usepackage[colorinlistoftodos]{todonotes}
\usepackage[colorlinks=true, allcolors=blue]{hyperref}
\usepackage{subfigure}
\usepackage[toc,page]{appendix} 
\usepackage{wrapfig}
\usepackage{aas_macros}
\usepackage{pifont}
\newcommand{\cmark}{\ding{51}}%

\newcommand{\done}{\rlap{$\square$}{\raisebox{2pt}{\large\hspace{1pt}\cmark}}}%

\begin{document}
\date{}

\title{\bf Astro2020 Science White Paper\\\vspace{0.2in} Characterizing Extra-solar Oort Clouds with Submillimeter-wave Observations}
\maketitle

\noindent \textbf{Thematic Areas:} \hspace*{60pt} $\done$ Planetary Systems \hspace*{10pt} $\done$ Star and Planet Formation \hspace*{20pt}\linebreak
$\square$ Formation and Evolution of Compact Objects \hspace*{20pt} $\square$ Cosmology and Fundamental Physics \linebreak
  $\square$  Stars and Stellar Evolution \hspace*{1pt} $\square$ Resolved Stellar Populations and their Environments \hspace*{20pt} \linebreak
  $\square$    Galaxy Evolution   \hspace*{45pt} $\square$             Multi-Messenger Astronomy and Astrophysics \hspace*{65pt} \linebreak

\noindent\textbf{Principal Authors:}
\\
Names: John Orlowski-Scherer, Eric Baxter, Cullen Blake, Mark Devlin, Bhuvnesh Jain \\
Institutions: University of Pennsylvania\\
Emails: \texttt{jorlo@sas.upenn.edu}, \texttt{ebax@sas.upenn.edu}\\

 \begin{abstract}
The Oort cloud, a collection of icy bodies orbiting the sun at roughly $10^{3}\,{\rm AU}$ to $10^{5}\,{\rm AU}$,  is believed to be the source of the long-period comets observed in the inner solar system.  Although its existence was predicted nearly 70 years ago, no direct detection of the Oort cloud has been made to date.  Given our current understanding of the formation of the Oort cloud, it is likely that many stars beyond the sun host their own {\it exo}-Oort clouds.  We have recently pointed out that submillimeter-wave telescopes have the capability to detect the thermal emission from these clouds.
High resolution observations of nearby stars with next generation submillimeter telescopes can obtain high significance detections of exo-Oort clouds, even in fairly conservative emission models.  A detection and characterization of such emission would open a new window into the study of the properties and evolution of planetary systems.
\end{abstract}   

\newpage \setcounter{page}{1}

\pagebreak
\section{Introduction}
\label{sec:introdution}
The Oort cloud is a hypothetical cloud of icy bodies believed to surround our solar system at distances from about $10^3$ to $10^5$ AU (see \cite{Dones:2004} and references therein for a review).  Oort originally posited the existence of the cloud to explain observations of long-period comets in the inner solar system, whose orbits are not confined to ecliptic plane \cite{Oort:1950}.  In current models, the cloud consists of a torus shaped inner cloud at some $10^3$ to $10^4$ AU from the sun, and a spherical outer cloud at roughly $10^4$ to $10^5$ AU.  

The standard picture for the formation of the Oort cloud begins with interactions between Jupiter and icy bodies in the early solar system, which should have resulted in a large population of bodies on highly elliptical orbits with large semi-major axes. Continued interactions with the giant planets would cause many of these bodies to become unbound, but interactions with nearby stars, the galactic potential, and nearby giant molecular clouds would circularize the orbits into a spherical cloud with semi-major axes up to $10^{5}\,{\rm AU}$ and random inclinations relative to the ecliptic plane.  Many questions remain about the precise mechanisms involved in the formation of the Oort cloud.   For instance, it has been argued that a significant fraction of the Oort cloud mass may be traced back to objects in planetary systems outside our own \citep{Levinson:2010}, and that the formation of the cloud may be connected to the properties of the stellar cluster in which the sun originated \citep{Brasser:2015}.

No direct detection of our own Oort cloud has been reported to date.  Detection via reflected light is difficult because of the large distances of the Oort objects from the sun, and because these objects likely have low albedo.  Detection by occultation of background objects is difficult because such events would be rare and have short duration \citep{Lehner:2016, Ofek:2010}.  One could also search for thermal emission from the Oort cloud.  The expected temperature of the Oort cloud objects is in the range of tens of Kelvin, leading to thermal emission in the submillimeter bands.  Detection of this emission from our own Oort cloud is challenging, largely because the signal is expected to be approximately isotropic on the sky, and is therefore difficult to separate from other backgrounds \cite{Babich:2007}. 

An alternative possibility is to detect the thermal emission from Oort clouds around {\it other} stars, which we refer to as exo-Oort clouds or EXOCs  \cite{Baxter:2018}.  One advantage of searching for an EXOC signal --- as opposed to that from our own Oort cloud --- is that the signal from an EXOC will be localized around its parent star, rather than uniformly distributed across the sky.  This spatial variation can be used to obtain an unambiguous identification of the signal by correlating submillimeter-wave maps of the sky with the positions of nearby stars. 

Detection of EXOCs around multiple stars would enable study of the connection between Oort cloud properties, and the properties of the stars and planetary systems that they surround, allowing Oort cloud formation and dynamical models to be tested.  Additionally, such measurements would constrain the properties of EXOC objects, which are relics of young planetary systems.  {\bf Detection of EXOCs would open a new window into the evolution and properties of planetary systems.}

The possibility of EXOC detection has been explored in IRAS data \cite{Stern:1991} and recently in {\it Planck} data \cite{Baxter:2018}.  These analyses placed limits on the properties of EXOCs that were close to the expected properties of our own Oort cloud.  Intriguingly, evidence for excess emission around a selection of hot, nearby stars was found in {\it Planck} data that appears broadly consistent with expectations for an EXOC \citep{Baxter:2018}.

Future submillimeter observations have the potential to robustly detect EXOC emission around large samples of stars.  One could imagine two types of EXOC searches: (1) targeted, high-resolution observations of single stars, and (2) wide-field surveys.  The advantage of the targeted observations is that greater map depths can be achieved.  The advantages of wide-field observations are that many stars can be observed, and that such observations are already planned (for different science cases) with future instruments.  {\bf Both targeted and wide-field approaches to EXOC detection are worth pursuing with future observations.}

\vspace{-0.1cm}

\section{The expected EXOC signal}
\label{sec:expect_signal}

The expected thermal emission signal from an EXOC is sensitive to its mass distribution, the size distribution of EXOC objects, the material properties of these objects, and the interaction of these objects with the surrounding radiation field. All of these properties are poorly constrained for our own Oort cloud.  {\bf However, even for conservative estimates of EXOC properties, a detection should be possible with future observations.}  

Mass estimates for our own Oort cloud are in the range of 5 to $20\,M_{\oplus}$ \cite{Dones:2004}, with some estimates being considerably higher.  This mass is expected to be concentrated in the inner parts of the cloud \cite{Howe:2014}.  Because of collisions, Oort cloud objects will fragment, resulting in a population of small grains and dust \cite{Pan:2005}.  The thermal signal from this population will be dominated by the smallest grains, making the signal sensitive to the small-scale cutoff in the size distribution \cite{Pan:2005}.  This cutoff is not well constrained, but estimates are in the range of about $1\, \mu {\rm m}$ to $10\, \mu {\rm m}$ \cite{Stern:1991,Howe:2014}.

The temperature of EXOC cloud objects is set by equilibrium between radiative input (from the parent star and the ambient radiation field), and thermal emission.  Since EXOC objects are likely mostly dirty ice \cite{Spinrad:1987}, we expect them to have low albedo of roughly $A \sim 0.03$.  Adopting the temperature and emissivity model from \cite{Stern:1991}, the temperature of the EXOC is expected to vary from about $10\,{\rm K}$ in the outer cloud to $30\,{\rm K}$ in the inner cloud for a sun-like star.  For a hotter star, the expected EXOC temperature is hotter.

\begin{figure*}
\centering 
  \includegraphics[width=0.65\textwidth]{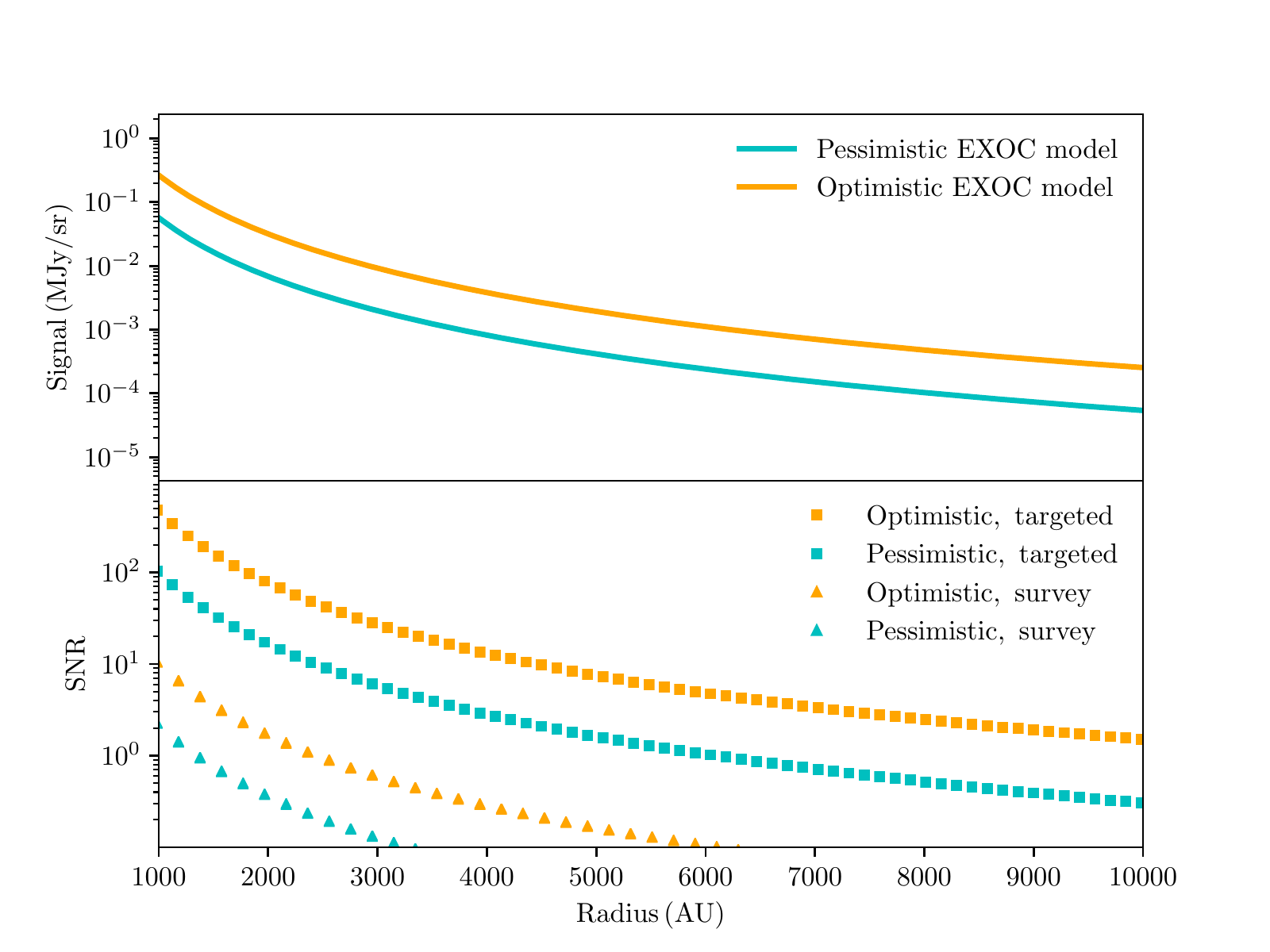}
\caption{\label{fig:projections}
\small Projected signal (top panel) and signal-to-noise ratio (bottom panel) for an EXOC measured around a nearby star.  We consider two EXOC models: one that yields a larger signal (optimistic), and one that yields a smaller signal (pessimistic).  Both models are allowed by current constraints.  We also consider two survey strategies: (a) targeted observations with high sensitivity, (b) survey-mode observations with lower sensitivity  (see details in \S\ref{sec:expect_signal}).   The points in the bottom panel indicate the choice of radial binning.  The signal-to-noise of the survey-mode observations could potentially be enhanced by averaging across multiple stars.}
\end{figure*}

As a case study, we estimate the expected signal for a possible EXOC surrounding 47 Ursae Majoris (47 UMa), which is at 14 pc, and is a promising candidate because it hosts a giant planet.  Combining the temperature model from \cite{Stern:1991} with the optical depth model from \cite{Baxter:2018}, the expected flux for 47 Uma is shown in the top panel of Fig.~\ref{fig:projections}.  We consider two model variations: one that yields a lower expected signal (pessimistic), and one that yields a larger signal (optimistic).  The pessimistic (optimistic) model assumes an EXOC mass of $5\,M_{\oplus}$ ($10\,M_{\oplus}$), a maximum radius of $5\times 10^4\,{\rm AU}$, and a minimum grain size of $10\,\mu{\rm m}$ ($3\,\mu{\rm m}$).   In both cases, we adopt the true distance and temperature of 47 UMa. 

In the bottom panel of Fig.~\ref{fig:projections} we show the expected signal-to-noise for different observational specifications.  We consider two projected sensitivity levels, corresponding to reasonable estimates for targeted and survey-mode observations.  For both observation modes, we assume an observation frequency of 900 GHz.  We assume beam sizes of 14$''$ (10$''$) and sensitivities of 0.1~MJy/sr (0.002~MJyr/sr) for the survey-mode (targeted) observations.  These sensitivities are reasonable for future observations, as we discuss in more detail below.  For the targeted observations, the outer parts of an EXOC can be probed at high signal-to-noise, even for the pessimistic model.  For survey-mode observations, the inner EXOC can be probed around individual stars.   The signal-to-noise of the survey-mode observations can be increased by averaging across many stars.  However, if only a small fraction of stars host Oort clouds, the  signal-to-noise of the star-averaged measurements will we reduced.

\section{Searching for Exo-Oort Clouds in the Next Decade}

Current generation instruments are close to being able to detect EXOCs around nearby, bright stars.  Next generation experiments will significantly improve the prospects of EXOC detection, and enable characterization of the radial profile of an EXOC signal, as well as determination of EXOC statistics across multiple stars.  

{\bf What types of observations are well suited to EXOC detection?}  Since the EXOC signal is expected to peak at around 900 to 1200~GHz (varying with EXOC properties and distance from the parent star), submillimeter observations in the range of about 200 to 1200~GHz are optimal for detection.  Observations beyond 1200~GHz would likely be in the Wien tail of the EXOC spectrum, and would require significantly better sensitivity.  High resolution observations are also important for resolving the radial profile of the inner EXOC.  With resolution of order $20^{\prime\prime}$, the inner radius of the outer Oort cloud should be resolvable out to 500~pc, beyond the height of the thick disk of the Milky Way.  In order to detect the outer parts of EXOCs, observations should extend to several tens of arcminutes around the parent stars.  An EXOC at 20~pc could subtend an angle of roughly 40 arcminutes (assuming a maximum size of $5\times 10^{4}\,{\rm AU}$).  High sensitivity observations are also needed to detect the small signal.  For the recommendations below, we assume that a several-$\sigma$ detection at 5000 AU is required.  To meet this requirement, searches for EXOC emission should have:
\begin{itemize}
  \setlength\itemsep{0.1em}
    \item {\bf Observation frequency}: ideally in the range 200 to 1200~GHz
    \item {\bf Resolution}: 0.5' or better resolution
    \item {\bf Map size}: 20' diameter or larger around each star
    \item {\bf Sensitivity}: 0.1 MJy/sr or better for observations averaged over hundreds to thousands of stars, 0.01 MJy/sr or better for observations of individual stars.
\end{itemize}

{\bf What instruments can enable EXOC detection?}  Several ongoing and near-term CMB surveys --- such as Advanced ACTPol \citep{Henderson:2016}, SPT-3G \citep{Benson:2014}, and the Simons Observatory \citep{SimonsObs2018} --- have the potential to place interesting constraints on EXOC properties in stacked analyses.  It is unlikely, though, that any of these current experiments could obtain high signal-to-noise detections of individual EXOCs, except perhaps in the very inner parts of the clouds.  The MUSTANG-2 instrument\footnote{\texttt{http://www.gb.nrao.edu/mustang/}} on the Greenbank Telescope has excellent resolution and sensitivity, although its frequency, 90GHz, is not optimal for EXOC observations. The Atacama Large Millimeter Array\footnote{\texttt{https://www.almaobservatory.org/en/home/}} (ALMA) has sufficient resolution, sensitivity, and frequency coverage to potentially enable EXOC detection. However, the maximum recoverable scale for the 870~GHz band on ALMA is only 7.7$''$, too small for the extended EXOC emission of a nearby star. 

Several longer term experiments will have the necessary wavelength coverage, resolution, and mapping speed to enable detection of EXOCs.  CCAT-prime will cover the relevant frequency range and has the necessary angular resolution (roughly 14$''$) to probe EXOCs.  The survey-mode sensitivity of 0.1~MJy/sr assumed in Fig.~\ref{fig:projections} is only a factor of a few better than the sensitivity of the planned Star Formation History survey with CCAT-prime.  The Probe of Inflation and Cosmic Origins (PICO) satellite would provide 800~GHz maps with nearly an order of magnitude lower noise than {\it Planck}, and resolution that is roughly in line with the requirements stated above \cite{PICO:2019}.   Additionally, the TolTec\footnote{\texttt{http://toltec.astro.umass.edu/}} instrument on the Large Millimeter Telescope will be capable of making maps with $1.3^{\prime \prime}$ resolution at 273~GHz, again fairly well matched to EXOC detection.  The Far-Infrared Imager and Polarimeter on the proposed Origins Space Telescope\footnote{\texttt{https://asd.gsfc.nasa.gov/firs/}} would also enable high signal-to-noise EXOC detection.

{\bf Why attempt an EXOC detection?}  The Oort cloud is one of the most poorly understood, and poorly constrained parts of our own solar system.  For this reason, a detection of EXOCs is an exciting prospect, with the potential to fill in serious gaps in our understanding of the properties and dynamics of planetary systems.  Submillimeter observations in the next decade have the potential to unlock this powerful tool, opening a new avenue for expanding our understanding of planetary science.

\clearpage
\bibliographystyle{plain}
\bibliography{ref.bib}

\end{document}